# Intrinsic Non-linearity of Josephson Junctions as an Alternative Origin of the Missing First Shapiro Step

Lei Xu[1,2,5], Shuhang Mai[3,5], Manzhang Xu[3,5], Xue Yang[1,2], Lihong Hu[1,2], Xinyi Zheng[1,2], Sicheng Zhou[1,2], Siyuan Zhou[1,2], Bingbing Tong[1,4], Xiaohui Song[1], Jie Shen[1,2,4], Zhaozheng Lyu[1,2], Ziwei Dou[1,2], Xiunian Jing[1], Fanming Qu[1,2,4], Peiling Li[1,2,4*], Guangtong Liu[1,2,4*], and Li Lu[1,2,4]

[1] *Institute of Physics and Beijing National Laboratory for Condensed Matter Physics, Chinese Academy of Sciences, Beijing 100190, China*
[2] *School of Physical Sciences, University of Chinese Academy of Sciences, Beijing 100049, China*
[3] *State Key Laboratory of Flexible Electronics (LoFE) & Institute of Flexible Electronics (IFE), Northwestern Polytechnical University, 127 West Youyi Road, Xi'an, 710072, China*
[4] *Hefei National Laboratory, Hefei, Anhui 230088, China*
[5] *These authors contributed equally: Lei Xu, Shuhang Mai, and Manzhang Xu*
\* *Corresponding authors:* cioran@iphy.ac.cn, gtliu@iphy.ac.cn

**Abstract**

The missing first Shapiro step in microwave-irradiated Josephson junctions has been widely interpreted as a hallmark of Majorana bound states. However, conventional mechanisms like junction underdamping or Joule heating can produce similar signatures. Here, we demonstrate that the intrinsic non-linear current-voltage characteristic of low-to-moderate transparency junctions can also suppress the first step, accompanied by distinctive zigzag boundaries between the zeroth and first step at intermediate driving frequencies. Microwave measurements on Al/WTe$_2$ junctions and numerical simulations of a non-linear resistively and capacitively shunted junction model reveal the first-step collapse induced by switching jumps of current, together with zigzag features absent in scenarios solely driven by finite $β$ or Joule heating. This zigzag signature therefore provides a crucial diagnostic tool, emphasizing the necessity of comprehensive analysis of microwave spectra before attributing the absence of the first Shapiro step to Majorana physics.

**Introduction**

Topologically protected fault-tolerant quantum computation relies on the non-Abelian statistics of Majorana zero modes (MZMs), which are Ising anyons predicted to exist in topological superconductors. Among the various experimental schemes proposed to detect MZMs, the fractional Josephson effect stands out as a crucial signature[1–3]. This effect arises from Majorana bound states (MBSs) within a topological Josephson junction, which can create a distinct current-phase relationship characterized by a 4π periodicity. The presence of this 4π periodicity can be verified via the observation of doubled Shapiro steps, as evidenced by reports in multiple experimental systems[4–12].

In the presence of radio frequency (RF) signals, phase locking induces a direct current (DC) component in the voltage-current characteristics (VICs) of a Josephson junction at voltage levels given by $V_n = nhf/2e, n \in Z$, where $h$ is Planck constant, $f$ is the RF frequency, and $e$ is the elementary charge. These voltage levels correspond to the so-called Shapiro steps. In a topological Josephson junction with fractional Josephson effect, only the steps corresponding to even values of n are expected to be maintained, yielding $V_n = nhf/e$. Consequently, under RF irradiation, the VICs of a topological junction should exhibit an absence of all odd-numbered Shapiro steps, resulting in the observation of doubled Shapiro steps. However, conclusive experimental evidence for the fractional Josephson effect remains elusive. While a few studies have reported the disappearance of multiple odd Shapiro steps[4], most observation are limited to the absence of the first step, a phenomenon often ascribed to voltage-dependent quasiparticle poisoning rates in non-trivial cases[13].

It should be noted that several distinct mechanisms unrelated to topology can account for the missing first Shapiro step. For instance, specific VICs can reproduce this feature, as demonstrated in previous studies[14–16]. Mudi *et al*. proposed a bias-dependent resistance model that predicts the absence of odd steps[14]. Meanwhile, Wu *et al*.[15] and Ustavschikov *et al*.[16] emphasized the role of large resistance near the switching current, manifested as steep switching jumps, which directly leads to the missing first step. Such switching jumps can stem from junction underdamping (a finite Stewart-McCumber constant $\beta$) or certain heating effects[16]. These frameworks can be applicable to both hysteretic and non-hysteretic junctions. For a non-hysteretic junction, it requires a suitable $\beta$ or heating to maintain the non-hysteretic VICs. However, for a hysteretic junction, a sufficiently large $\beta$ or substantial heating effects can induce hysteresis along with switching jumps.

In this work, we fabricate Josephson junctions based on WTe$_2$ nanoflakes and observe the absence of the first Shapiro step. Detailed analysis of the VICs reveals pronounced switching jumps with minimal hysteresis, which previously attributed to a finite $\beta \gtrsim 1$ or heating. Interestingly, in addition to the non-hysteretic VICs, we simultaneously observe an intriguing feature in the Shapiro spectra—zigzag transition lines at intermediate RF frequencies (~ 3 – 5 GHz). We propose that this distinct behavior can be effectively described by employing a non-linear resistively and capacitively shunted junction (NRCSJ) approach[17]. Unlike the standard RCSJ model, which assumes a constant shunting resistance, the NRCSJ approach incorporates a voltage-dependent resistance—an intrinsic property of real Josephson junctions often overlooked in conventional treatments. Within the NRCSJ framework, the observed switching jumps do not necessitate a finite $\beta \gtrsim 1$ or certain heating effects, allowing for the universal reproduction of notable characteristics, including the zigzag transition lines and the missing first Shapiro step, owing to the inherent non-linearity of the system.

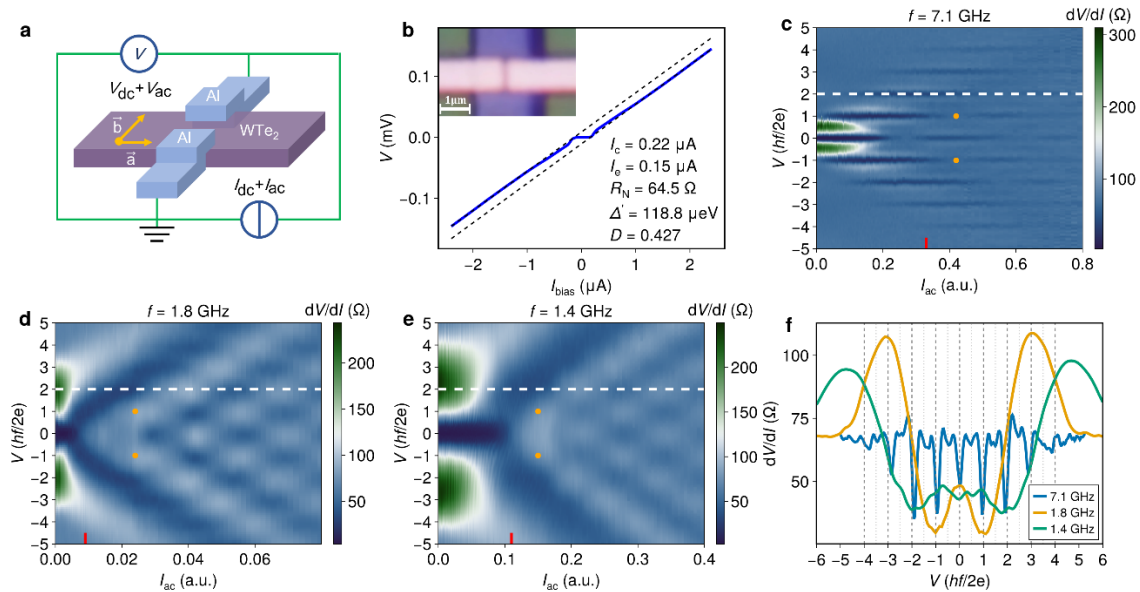

**Fig. 1 Electrical characterization of a WTe₂-based Josephson junction. a** Schematic of the Josephson junction device. The orange axes indicate the crystallographic orientation of the WTe₂ nanoflake. **b** Voltage-current (*V-I*) curves of the junction. The dashed lines are linear asymptotes of the high-bias behavior. Inset: an optical image of a typical device. **c-e** Differential resistance (d$V$/d$I$) maps as a function of AC current (arbitrary units) and normalized voltage (scaled to $V_0 = hf/2e$) under different RF driving frequencies. White dashed lines are guides to the eye highlighting the position of the second Shapiro step. Orange dots mark the critical points $P_c$ for the onset of the Bessel oscillation regime. **f** Vertical linecuts taken from the position marked by the red segments in panels (c)-(e). Dark and light dashed lines indicate the positions of integer and half-integer Shapiro steps, respectively.

## Results

To conduct microwave irradiation experiments, we fabricated multiple planar Josephson junctions based on WTe$_2$. Fig. 1a shows the typical device layout. The WTe$_2$ ultra-thin flakes used for the Josephson junctions were grown by chemical vapor deposition (CVD) method[18]. Superconducting Ti/Al electrodes were deposited directly on the surface of WTe$_2$ using electron-beam evaporation, forming a junction with dimensions of $L \sim 100$ nm in length and $W \sim 1$ μm in width. Fig. 1b shows the VICs of the junction. The main panel reveals a critical current $I_c$ is ~ 0.22 μA. By fitting the high-bias asymptotic, we extract the excess current $I_e \sim 0.15$ μA and normal state resistance $R_N \sim 64.5$ Ω. The induced gap $\Delta' \sim 118.8$ μeV can be determined from voltage bias results shown in Fig. S1. Based on the Ref.[19], the transparency of the junction $D \sim 0.427$ can be derived with the above parameters (see Supplementary Note 1). Moreover, significant switching jumps can be observed, which are often ascribed to a finite $\beta$[15], heating effects as suggested in previous studies[16], or intrinsic non-linear effects, which will be discussed in detail later.

To probe the Shapiro steps, we irradiated the sample with microwaves via a suspended RF antenna. Figs. 1c-e display the spectra of differential resistance $dV/dI$ as a function of AC current (in arbitrary units derived from RF power with 50 Ω impedance) and voltages normalized by $hf/2e$. In these spectra, Shapiro steps are manifested by dark blue stripes. At a frequency of $f = 7.1$ GHz (Fig. 1c), all integer Shapiro steps are visible across the measured power range, accompanied by faint half-integer Shapiro steps. These fractional steps may arise from high-transparency conduction channels or chaotic dynamics induced by $\beta$, as demonstrated by our numerical simulations (Fig. 4d). These behaviors are consistent with typical Shapiro response in SNS junctions[20]. As the driving frequency decreases to 1.8 GHz, the first Shapiro step almost disappears in the low power range, as illustrated in Fig. 1d, but recovers at higher driving amplitudes. At 1.4 GHz, the first step is completely absent for $I_{ac} < 0.18$, though higher-order steps remain visible (Fig. 1e). Fig. 1f plots three representative linecuts taken at the positions marked by the red segments in Figs. 1c-e. The dips corresponding to the first and negative first Shapiro steps ($V_n = \pm 1$) become progressively rounded and eventually disappear as the frequency decreases. The half-integer Shapiro steps, indicated by lighter dotted lines, appear as small dips in the 7.1 GHz trace. These linecuts profiles clearly illustrate the progressive suppression of the first Shapiro step and the emergence of half-integer steps at higher frequencies.

The absence of the first Shapiro step is often taken as a signature of a 4π-periodic supercurrent. However, such a topological interpretation requires careful scrutiny in the context of our WTe$_2$-based SNS junctions. At low temperatures, WTe$_2$ can exist in multiple distinct phases: a two-dimensional topological insulator for the monolayer case[21], and a higher-order topological insulator[22] or Weyl semimetal[23] for thicker flakes. Theoretical works have predicted that MZMs emerging in these topologically non-trivial systems could indeed give rise to a fractional (4π-periodic) Josephson effect[3,24,25]. Nevertheless, several aspects of our experimental design can rule out this interpretation. In our WTe$_2$-based devices, the superconducting electrodes are intentionally patterned to avoid contacting the sample edges, thereby minimizing contributions from topological edge states. Furthermore, for Fermi-arc surface states related to the Weyl semimetal phase, Ref.[26] shows that the fermion-parity is not preserved, resulting in a conventional 2π-periodic current-phase relationship despite the gapless spectrum. While several works[27,28] have suggested that Landau-Zener transition (LZT) could effectively restore a 4π periodicity, time-reversal symmetry forces the gapless Andreev bound states to merge with the continuum. This mechanism significantly enhances quasiparticle poisoning rates, which likely suppress any observable 4π periodicity. Given the multiplicity of possible electronic phases and mechanisms, it is challenging to attribute the observed missing first Shapiro step to a 4π supercurrent. As we demonstrate below, a trivial explanation could account for these observations.

To further elucidate this behavior, we denote the critical point of marking the onset of oscillations in the first Shapiro step patterns as $P_c$, represented by the orange points in Fig. 1c-e. Interestingly, this sharp threshold corresponds precisely to the maximum power at which the first Shapiro step is absent at the lowest frequency of $f = 1.4$ GHz shown in Fig.1e. Consistent with previous reports, once the applied microwave power exceeds $P_c$, all Shapiro steps reappear and exhibit conventional 2π behavior. Prior works[6–11] attributed this phenomenon to the emergence of a 4π supercurrent carried by MBSs. In contrast, models based on the bias-dependent quasiparticle-poisoning rates do not support this special threshold[13]. However, these models merely qualitatively suggest that increased poisoning rates at higher voltages destroy the 4π periodicity, they failed to identify the critical poisoning rate required for the suppression. This means that the onset of the absence of 4π periodicity can occur at any voltage, rather than being strictly correlated with the first Bessel node. Recent investigations have suggested that switching jumps can also contribute to this unique phenomenon[15,16]. Moreover, corresponding simulations suggest that the missing first Shapiro step can only occur below the critical point $P_c$, in line with the experimental observations. By introducing a finite $\beta \gtrsim 1$, this theory can be used to explain part of our results (Figs. 1b-f). However, detailed experimental results of the frequency-dependent spectral evolution (see Supplementary Fig. S3) reveal that existing theoretical proposals, including those incorporating switching dynamics and finite $\beta$, cannot fully capture the observed evolution of the Shapiro steps. These discrepancies highlight the need for modifications to existing theoretical models.

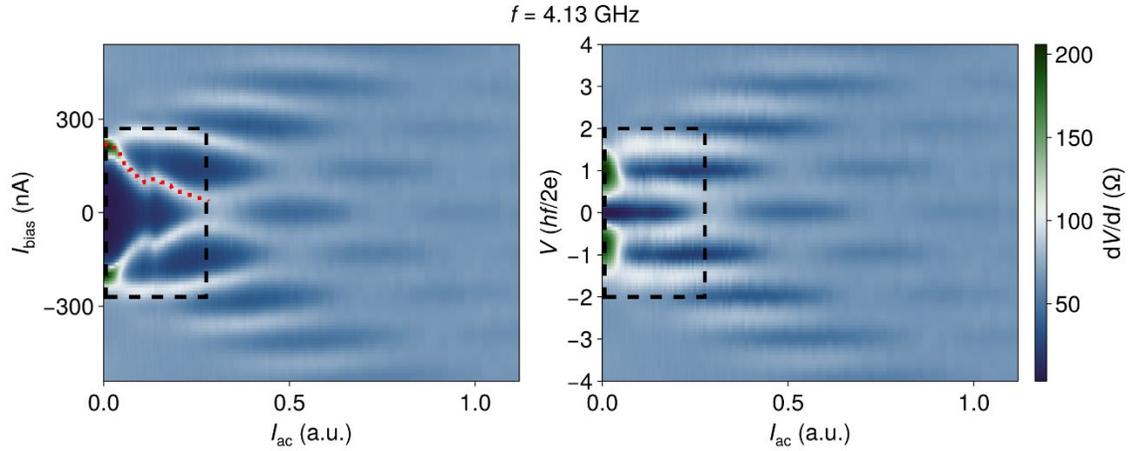

**Fig. 2 Zigzag transition line in the Shapiro step structure. a** Differential resistance ($dV/dI$) map as a function of DC bias current ($I_{bias}$) and AC-drive amplitude ($I_{ac}$). The black dashed box highlights the zigzag transition region, while the red dashed line traces its positive branch. **b** Corresponding voltage–current characteristics (VICs) obtained by numerical integration of the data in panel (a). The black dashed box denotes the region corresponding to the zigzag region shown in panel (a).

Fig. 2a shows the differential resistance spectrum at a moderate frequency $f = 4.13$ GHz. Overall, the observed features are consistent with a conventional Josephson junction under RF irradiation. However, a distinct zigzag transition line, marked by a kink at its midpoint, is observed at the boundary between the zeroth and first Shapiro steps. To determine whether this special transition line influences the overall Shapiro step structure, we present the corresponding integral plot in Fig. 2b. In this spectrum, the zigzag pattern becomes indistinguishable from the background features. The anomalous feature is highlighted by a black dashed box in Figs. 2a and 2b. It is noteworthy that the zigzag feature is more prominent in current-bias experiments than in voltage-bias measurements. Moreover, Supplementary Fig. S3 reveals that this phenomenon occurs exclusively within a specific frequency range, accompanied by a frequency-dependent evolution of the kink's position. Specifically, as the frequency increases, the kink initially emerges at higher power levels before the first oscillation node, then shifts to lower power and eventually disappears.

This anomalous zigzag transition line and its evolution cannot be explained by the conventional RCSJ model (as demonstrated in Supplementary Note 5), motivating us to introduce an extended framework to account for the experimental observations. Therefore, we incorporate a voltage-dependent resistance $R_N(V)$ into the RCSJ model, as illustrated in Fig. 3a. In dimensionless units, the resulting model is described by,

$$i + i_{ac}\cos(\omega_{ac}\tau) = \beta\ddot{\varphi} + i_N(v) + i_e(v) + i_s(\varphi) \tag{1}$$

where $i = I/I_c$, $i_{ac} = I_{ac}/I_c$, $V_J = I_c R_N$, $v = V/V_J$, $\omega_c = 2eV_J/\hbar$, $\tau = \omega_c t$, $\omega_{ac} = \Omega_{ac}/\omega_c$, $\beta = \omega_c R_N C$, $i_N = I_N/I_c$, and $i_s = I_s/I_c$. The term $i_e(v) = \alpha_1 \tanh(\alpha_2 v)$ accounts for the excess current[29]. Unlike the standard RCSJ model, which assumes a linear normal current, $i_N(v) = v$, our proposed NRCSJ model inspired by Ref.[17] introduces a new form for the normal current,

$$i_N(v) = \frac{v}{1 + \left(\frac{v_c}{v}\right)^{2n}}, n \in Z, n \geq 1 \tag{2}$$

where $v_c = V_c/V_J$ and n are two free parameters, which characterize the critical voltage and the intrinsic non-linearity of a conventional Josephson junction (see Supplementary Note 3 for derivation details). To illustrate the characteristics of this model, we plot the corresponding current-voltage characteristics (IVCs) in Figs. 3b and 3c. In Fig. 3b, we fix n = 18 and examine the evolution of the IVCs as $v_c$ increases from 0 to 2.0. For $v_c = 0$, the curve recovers the linear behavior of the standard RCSJ model. For $v_c \neq 0$, the corresponding current $i$ exhibits a steep jump around $v_c$, beyond which the curve asymptotically approaches the $v_c = 0$ case. This indicates that $v_c$ sets the threshold voltage for the sharp transition. To further elucidate the role of the index n, Fig. 3c shows the evolution of IVCs as a function of the non-linearity order n at fixed $v_c = 1$. As n increases from 2 to 32, the sharpness and the onset of the transition near $v_c$ become more pronounced, demonstrating that n controls the non-linearity of the response. Thus, while $v_c$ determines the position of current jump, n modulates both the sharpness and the onset of the transition. With these two parameters, the NRCSJ model captures the non-linear behavior of most junctions with low-to-moderate transparencies, as is the case for our devices (see Supplementary Note 3 for more details).

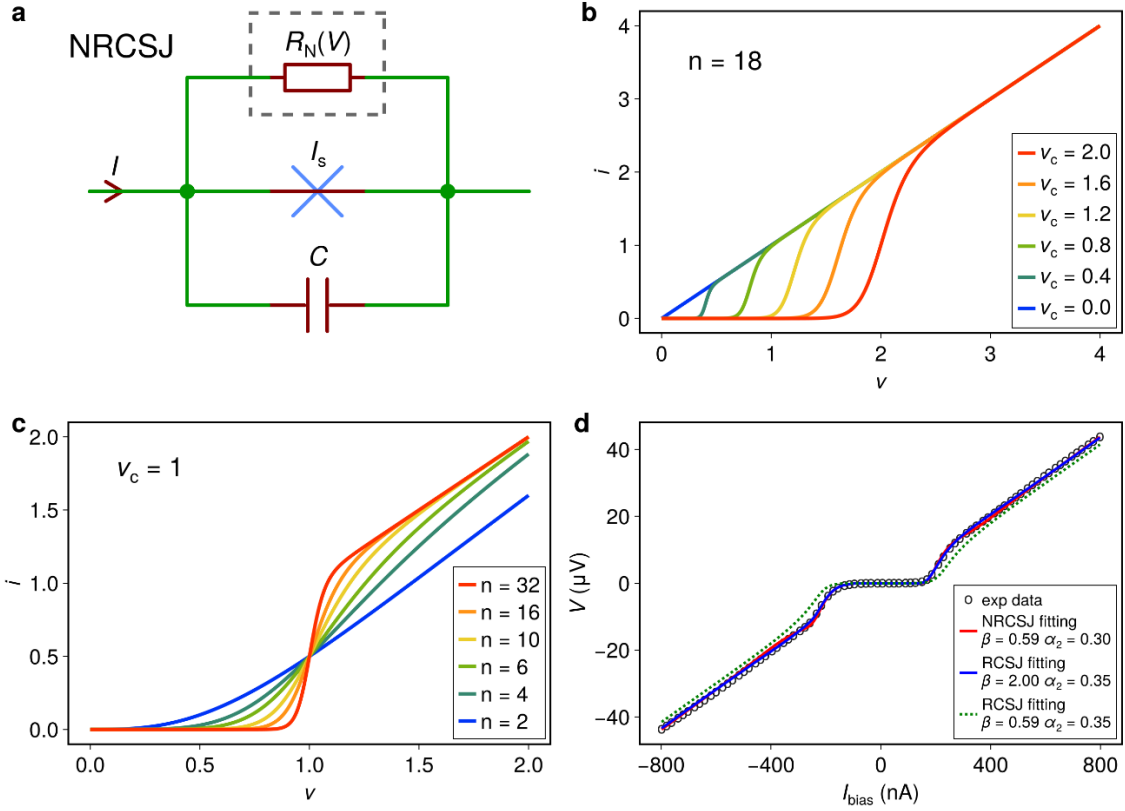

**Fig. 3 Non-linear RCSJ (NRCSJ) model and current-voltage characteristics (IVCs). a** Circuit schematic of the NRCSJ model. The gray-dashed box highlights the proposed voltage-dependent non-linear normal resistance, which replaces the linear resistor in the conventional RCSJ model. **b** Non-linear normal current-voltage (*I-V*) characteristics under different critical voltages $v_c$ at fixed n=18. **c** Evolution of the current–voltage curves (*I-V*) as a function of the non-linearity order n at fixed $v_c$ =1. **d** Comparison of experimental *V-I* data with fits from the proposed NRCSJ model and the conventional RCSJ model. The NRCSJ fitting was obtained with $v_c = 0.64$ and n = 20.

To evaluate the effectiveness of the proposed model, Fig. 3d compares the measured VICs with simulations generated using both the conventional RCSJ model and the proposed NRCSJ model. To reproduce the experimental data accurately, both the excess current term $i_e(v)$ and smoothing effects from lock-in amplifier have been considered (See Supplementary Note 2). The experimental VICs exhibit a steep jump when the junction switches from superconducting state to the normal state, with negligible hysteresis, a feature known to arise from an appropriate $\beta$. Optimal fitting yields $\beta = 2.0$ for the RCSJ model and $\beta = 0.59$ for the NRCSJ model, both of which show satisfactory agreement with the data. It should be noted that, as the dashed green curve in Fig. 3d demonstrates, the conventional RCSJ model with $\beta = 0.59$ fails to produce a switching jump of the observed sharpness, in which $\beta \sim 1$ is the boundary for overdamping and underdamping cases. This discrepancy highlights a fundamental difference in the physical dynamics captured by the two models.

The emergence of switching jumps in the NRCSJ model at a much smaller $\beta$ compared to the RCSJ model can be understood from two complementary perspectives. First, once the current bias slightly exceeds the critical current of the Josephson junction, a finite voltage jump $v_c$ is required by the non-linear IVCs for the portion of current beyond $I_c$ to flow through the shunt resistance. This contrasts with the linear IVC case, where $v_c = 0$. Second, the dynamics of the junction are governed by an effective $\beta$ that becomes large near the switching current. Although a conventional large $\beta$ would produce steep switching jumps akin to the RCSJ model, the non-linear IVC here leads to a spike in $R_N(V)$ around the switching current. This results in a large $\beta$ in the vicinity of $I_c$, while the junction remains overdamped ($\beta \ll 1$) at other current bias values. This characteristic can induce sharp switching jumps without hysteresis, mimicking an underdamped system only at the transition.

Despite both models yielding acceptable results in Fig. 3d, the unique zigzag feature observed in the RF irradiated spectra (Fig. 2a) —unreproducible by the conventional RCSJ model—provides critical validation for the proposed NRCSJ model. With parameters extracted from Fig. 3d, we simulated the Shapiro response under microwave irradiation using the NRCSJ model. As shown in Fig. 4a, the simulation successfully reproduces the zigzag transition lines between the zeroth and the first steps observed experimentally. To evaluate the simulation results, we extracted the zigzag trajectories from Fig. 2a and Fig. 4a and plotted them in Fig. 4b. A good agreement between the experimental data and simulated results can be resolved. Furthermore, the simulated frequency evolution (Supplementary Fig. S3 and Fig. S5) shows overall consistency with the data. In contrast, the RCSJ model with the same parameters fails to generate any zigzag feature at any frequency (Supplementary Fig. S4).

Further evidence of the model's validity is shown in Fig. 4c. With decreasing the driving frequency, the first Shapiro step is gradually suppressed and finally disappears at driving powers below $P_c$, coinciding with the first node entering the Bessel oscillation regime. Moreover, as shown in Fig. 4d, the NRCSJ model accurately reproduces the half-integer Shapiro steps observed at higher frequencies. This simulation suggests that the half-integer steps stem from a synergistic effect of finite junction transparency, a specific damping parameter $\beta$, and circuit non-linearity. Therefore, even without topological nontrivial contribution, the NRCSJ model can capture all key features of the RF irradiation spectra in WTe$_2$ Josephson junctions, including the zigzag transitions and the suppression of the first Shapiro step. This demonstrates that the missing first Shapiro step, often considered as preliminary evidence for $4\pi$-periodic supercurrents caused by MBSs, can be explained by the non-linear dynamics of a conventional junction.

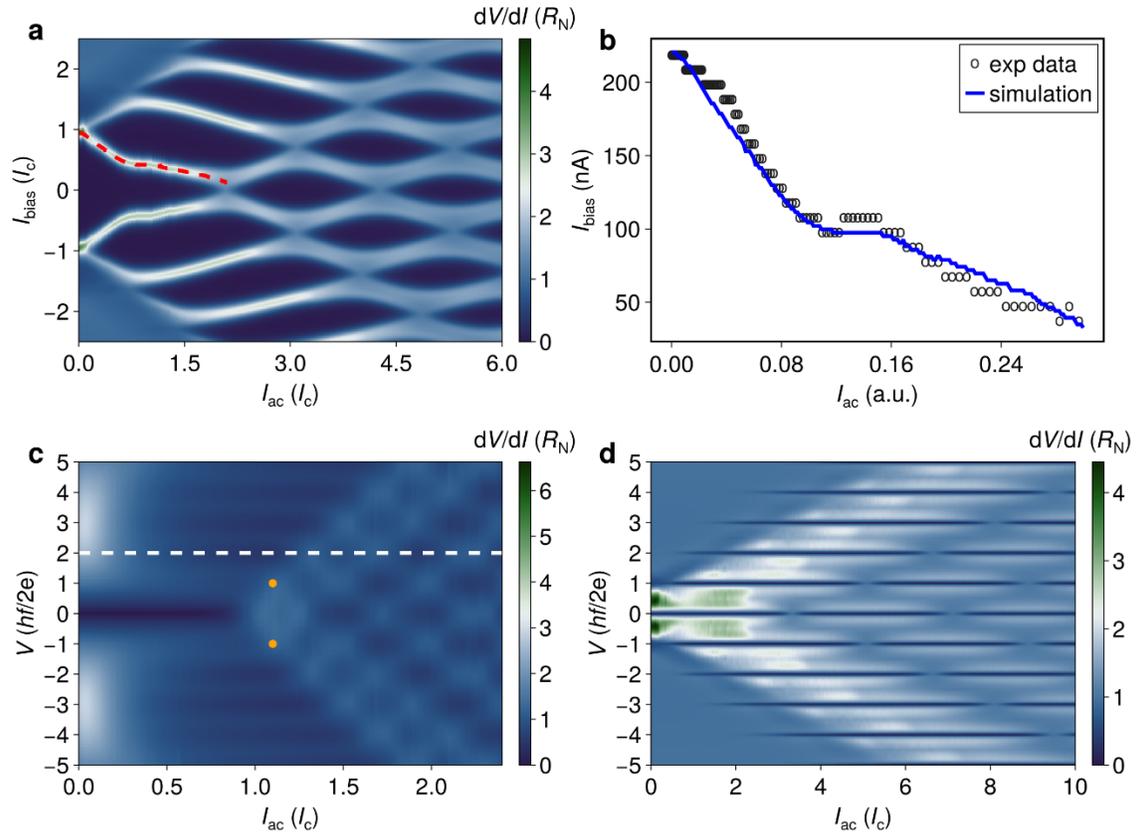

**Fig. 4 Simulated spectra based on the NRCSJ model. a** Calculated microwave-irradiated spectrum exhibiting the characteristic zigzag transition lines. The red dashed line traces the positive-voltage branch of zigzag transition lines. **b** Superposition of the zigzag transition lines extracted from the experimental map in Fig. 2a and the simulated map in Fig. 4a. **c** Simulated low-frequency spectrum that corresponds to Fig. 1e, demonstrating the missing of the first Shapiro step. White dashed line is guide to the eye highlighting the position of the second Shapiro step. Orange dots mark the critical points $P_c$ for the onset of the Bessel oscillation regime. **d** Simulated high-frequency spectrum corresponds to Fig. 1c, exhibiting pronounced half-integer Shapiro steps.

## Discussion

The absence of odd Shapiro steps has been proposed as a key signature of MZMs.

However, this feature can also arise from several trivial mechanisms, which can be classified into two main categories. The first includes mechanisms such as LZT[28,30,31], Leggett modes[32], axion field[33], and specialized measurement circuit[34], all of which can lead to 4π-like phenomena. The second category involves mechanisms like switching jumps in the VICs, often attributed to a certain $\beta$ or heating effects[16]. A crucial distinction is that the latter can only account for the missing first Shapiro step, while the former can induce suppression of higher odd-order steps. Based on our experimental results, we can rule out all mechanisms from the first category. Effective LZT requires high junction transparency, a condition not fulfilled by our junction, which has a relatively low transparency of $D \approx 0.427$. Leggett modes require multiband superconductivity and are restricted to a narrow frequency range, while the axion field is specific to axion insulator systems. Since our WTe$_2$-based junction meets none of these criteria, these mechanisms are inapplicable. The scenario of a specialized circuit configuration is also eliminated, as it requires considerable shunting inductance and capacitance. The minimal hysteresis observed in our VICs suggests that the contributions from inductance and capacitance are negligible. Therefore, the missing first Shapiro step in our system is likely associated with the occurrence of switching jumps in the VICs. While previous reports[15,16] have attributed such jumps to a finite $\beta$ or heating effects, these factors alone cannot fully explain our experimental results within the standard RCSJ model.

Our proposed NRSCJ model successfully accounts for all key experimental observations, including the switching jumps, the concomitant missing first Shapiro step, and the distinct zigzag transition lines in the microwave spectra. Furthermore, by analyzing the physical origin of the non-linear effects, we find that such behavior is widely present in Josephson junctions (see Supplementary Note 3), as captured by the effective non-linear current-voltage relationship given in Eq. (2). Considering a single-channel SNS junction with different transparencies, the theoretical IVCs derived from the BdG equation[35] exhibit non-linearity with three key features (Supplementary Fig. S2a): 1) a nearly vanishing normal current at low voltages, 2) an excess current at high voltages, and 3) signatures of multiple Andreev reflections (MARs) at voltages below twice the superconducting gap. Our model primarily incorporates the first characteristic through the nonlinear form of $i_\text{N}(v)$. In the tunnel limit, setting $V_\text{c} = 2\Delta$ and a large index n yields excellent agreement with theoretical curve in Supplementary Fig. S2b, both exhibiting a steep current jump. In junctions with low-to-medium transparency, the transition voltage $V_\text{c}$ becomes smaller and the switching behavior becomes smoother, consistent with the cases shown in Figs. 3b and 3c. The second characteristic—the excess current—is well described by an existing effective model[29] and included into Eq. (1) via the term $i_e(v)$. The third characteristic, prominent in single-channel junctions[35], is not observed in our multi-channel WTe$_2$-based junctions, where contributions from both bulk and surface states result in a smooth IVC without pronounced MAR peaks. Hence, the effective non-linear resistance described by Eq. (2) is sufficient to capture the non-linear behavior of our SNS junctions, and indeed provides satisfactory simulations of the experimental microwave spectra. Above analysis suggests that junctions with low-to-medium transparency intrinsically possess such non-linearity, implying that the switching jumps may exist in a wide range of SNS junctions. Consequently, the missing first Shapiro step could also be a widespread phenomenon, provided that the relative driving frequency $\omega_{ac}$ is sufficiently low, as demonstrated by the simulations in Fig. 4c.

In summary, we fabricated Josephson junctions using the Weyl semimetal WTe$_2$ and investigated their response under microwave irradiation. Our measurements reveal a significant deviation from the standard RCSJ model. Specifically, in the mediate frequency regime, we observe pronounced zigzag patterns in the transition between the zeroth and first Shapiro steps. These observations underscore the critical influence of junction non-linearity and highlight the need of comprehensive frequency-dependent measurements to discriminate among alternative origins of the missing first Shapiro step and uncover the underlying mechanisms.

Motivated by these findings, we developed a NRCSJ model that incorporates an effective voltage-dependent normal resistance. This model successfully accounts for both the missing first Shapiro step and the distinctive zigzag transition lines observed in our experiments. Our analysis establishes that intrinsic non-linearity is a key driver for the occurrence of switching jumps in the VICs, which in turn lead to the missing first Shapiro step. Consequently, these findings necessitate a critical re-examination of the missing first Shapiro step as definitive evidence of 4π-periodic supercurrent resulting from MBSs.

**Methods**

WTe$_2$ thin flakes were grown on a Si/SiO$_2$ substrate by chemical vapor deposition. To protect the samples, a spin-coated electron beam photoresist (PMMA A5) was applied. Prior to device fabrication, the photoresist was removed by soaking in acetone for several minutes, followed by rinsing in isopropyl alcohol. The samples were then transferred into a glove box to prevent exposure to the atmosphere. For device processing, PMMA A4 was spin-coated at 3000 rpm and baked at 120 °C for 2 minutes. The electrode patterns were defined using 20 kV electron beam lithography. After 1-minute development, the samples were etched by Ar ion milling for 1 minute, followed by in-situ deposition of Ti/Al (5/60 nm) to form superconducting electrodes.

Transport measurements were performed in an Oxford Triton dilution refrigerator with a base temperature of 10 mK. The voltage-current characteristics were recorded using LI5640 lock-in amplifiers and a Keithley 2612 source meter. Microwave irradiation was applied via an antenna suspended above the device. The microwave signals were generated using an Agilent E4421B (250 kHz – 3 GHz) and an E8254A (250 kHz – 40 GHz).

**Acknowledgment**

This work is supported by the National Basic Research Program of China from the Ministry of Science and Technology (MOST) under grants No. 2022YFA1602803 and 2023YFA1607400; by the National Natural Science Foundation of China under grants No. 12574060, 62304182, and 12104498; and by the Strategic Priority Research Program of the Chinese Academy of Sciences under grant No. XDB33010300. Support was also provided by the Synergetic Extreme Condition User Facility, and the Innovation Program for Quantum Science and Technology under Grant No. 2021ZD0302600.

**Data availability**
The data that support the plots within this paper and other findings of this study
are available from the corresponding author upon reasonable request.

**Author contributions**
Lei Xu performed the device fabrication, low-temperature transport measurements, data analysis, and carried out the final simulation-based interpretation. Shuhang Mai and Manzhang Xu prepared the WTe$_2$ samples. Xue Yang, Lihong Hu, Xinyi Zheng, Sicheng Zhou, and Siyuan Zhou assisted in the fabrication process. Bingbing Tong, Xiaohui Song, Jie Shen, Zhaozheng Lyu, Ziwei Dou, Xiunian Jing, Fanming Qu, and Li Lu provided measurement facilities and technical support. Guangtong Liu and Peiling Li supervised the project, contributed to experimental planning, provided resources, and revised the manuscript. All authors discussed the results and approved the final version of the manuscript.

**Competing interests**